\documentclass[prd,amsmath,aps,floats,amssymb, floatfix,
  superscriptaddress,nofootinbib,twocolumn,preprintnumbers]{revtex4-2}
\usepackage{tabularx}
\usepackage{amsmath,amssymb,natbib,latexsym,times}

\usepackage{booktabs}
\usepackage{multirow}

\usepackage{graphicx}
\usepackage{dcolumn}
\usepackage{bm}

\usepackage{url}
\usepackage{color}
\usepackage{comment}
\usepackage{mathrsfs}
\usepackage{hyperref}
\usepackage[dvipsnames]{xcolor}
\usepackage{orcidlink}
\usepackage{mathtools}
\usepackage{physics}
\AtBeginDocument{
  \renewcommand{\selectlanguage}[1]{\relax}
}
\usepackage{siunitx}

\newcommand{\B}[1]{{\boldsymbol{#1}}}

\newcommand{\mr}{\mathrm}

\newcommand{\sqbra}[1]{
\left[
#1
\right]
}
\newcommand{\parbra}[1]{
\left(
#1
\right)
}

\newcommand{\avrg}[1]{\left\langle #1 
\right\rangle}

\begin{document}
\title[]{
An optimal quadratic estimator for window-free cosmic shear power spectra
}

\author{Taisei~Terawaki\orcidlink{0009-0001-4213-7924}}
\email{taisei.terawaki@ipmu.jp}
\affiliation{Kavli Institute for the Physics and Mathematics of the Universe (WPI), The University of Tokyo Institutes for Advanced Study (UTIAS), The University of Tokyo, Chiba 277-8583, Japan}
\affiliation{Department of Physics, The University of Tokyo, Bunkyo, Tokyo 113-0031, Japan}

\author{Masahiro~Takada\orcidlink{0000-0002-5578-6472}}
\affiliation{Kavli Institute for the Physics and Mathematics of the Universe (WPI), The University of Tokyo Institutes for Advanced Study (UTIAS), The University of Tokyo, Chiba 277-8583, Japan}
\affiliation{%
Center for Data-Driven Discovery (CD3), Kavli IPMU (WPI), UTIAS, The University of Tokyo, Kashiwa, Chiba 277-8583, Japan}%

\begin{abstract}
The pseudo-$C_\ell$ estimator recovers the true cosmic shear power spectrum by correcting for the survey window convolution while employing inverse-variance weighting based on intrinsic shape noise of source galaxies. 
However, this weighting scheme is optimal only on small angular scales where shape noise dominates.
In this paper, we derive a quadratic estimator for the unwindowed cosmic shear power spectrum by maximizing the Gaussian likelihood of the pixelized galaxy-shape field using the full covariance matrix, which accounts for both sample variance and shape noise.
By combining FFTs in the flat-sky approximation, the conjugate-gradient method, and  Monte Carlo realizations of Gaussian ancillary fields, we substantially reduce the computational cost of estimating the Fisher matrix, a key ingredient of the estimator that requires repeated inverse-covariance matrix operations.
Using Gaussian simulations of shape fields, we validate the method and demonstrate that it can recover the input $E$-mode power spectrum with statistically optimal precision across all angular scales.
We then apply the method to shape fields generated from ray-tracing simulations for a $\Lambda$CDM cosmology and show that, compared with the pseudo-$C_\ell$ method, it reduces the statistical uncertainties in the $E$-mode power spectrum by 5--15\% at multipoles of $\ell \lesssim 500$.
We further demonstrate that the method significantly suppresses $E$- to $B$-mode leakage across the full multipole range. 
Our estimator therefore provides a statistically optimal approach for measuring cosmic shear power spectra from wide-area galaxy survey data.
\end{abstract}

\maketitle

\section{Introduction}
\label{sec:intro}

Weak gravitational lensing (hereafter referred to as cosmic shear) has emerged as one of the most powerful probes of cosmology~\citep{2018ARA&A..56..393M}.
Because it directly probes the underlying matter distribution without being affected by uncertainties in galaxy bias, it enables precise constraints on key cosmological parameters, particularly the matter density $\Omega_{\rm m}$ and the amplitude of matter fluctuations $\sigma_8$~\citep[e.g., see Refs.][for recent results]{2023PhRvD.108l3518L,dalal_hyper_2023,2025A&A...703A.158W,DES_Y6_cosmicshear}.

The statistic most commonly used in cosmological analyses of cosmic shear is the two-point correlation function (2PCF) of galaxy shapes~\citep{1991MNRAS.251..600B,1991ApJ...380....1M,1992ApJ...388..272K}.
The 2PCF has both advantages and disadvantages. Its main advantage is that it is unaffected by the survey window function. 
On the other hand, a major drawback is that 2PCF measurements in different separation bins are strongly correlated, leading to significant off-diagonal elements in the covariance matrix~\citep{2002A&A...396....1S,2011ApJ...734...76S,2019MNRAS.486...52S}. 
In addition, the $E$/$B$-mode decomposition of the 2PCF is non-local and is therefore not straightforward~\cite{2002A&A...389..729S}.

The Fourier-space counterpart of the 2PCF is the power spectrum~\citep{hu_power_2001,brown_shear_2003,kohlinger_kids-450_2017}. 
However, compared with 2PCF analyses, cosmological analyses based on the power spectrum  remain relatively underexplored~\citep[see Refs.][for representative examples]{brown_shear_2003,kohlinger_kids-450_2017,hikage_cosmology_2019,doux_dark_2022,dalal_hyper_2023}.
A naively estimated power spectrum is affected by the convolution with the survey window function~\citep{2013PhRvD..87l3504T}. 
In contrast, the covariance matrix of the power spectrum is more diagonal than that of the 2PCF, making it relatively straightforward to model~\citep{2009ApJ...701..945S,2009MNRAS.395.2065T,2013PhRvD..87l3504T,2017MNRAS.470.2100K}. 
Furthermore, the $E$/$B$-mode decomposition can be performed locally~\citep{stebbins_weak_1996,2002A&A...389..729S,2002ApJ...568...20C}.
Real-space (2PCF) and Fourier-space (power spectrum) cosmological analyses are theoretically equivalent when all angular scales are included. In practice, however, this equivalence generally breaks down because the range of scales used in the analysis is restricted by scale cuts and other data-selection criteria,
and because different weighting schemes are generally employed. Therefore, it is important to analyze the same survey data in both real and Fourier space and compare the resulting cosmological constraints~\citep[see, e.g., Refs.][for related discussions]{2020PASJ...72...16H,2025PhRvD.111f3509T}.

A standard method for estimating the underlying cosmic shear power spectrum while correcting for the effects of the survey window function is the pseudo-$C_\ell$ estimator~\citep{brown_shear_2003,hikage_pseudo-spectrum_2016},
which was originally developed for CMB analysis~\citep[e.g.,][]{wandelt_pseudo-c_l_2001}. 
A widely used public implementation of this estimator is \texttt{NaMaster}~\citep{alonso_unified_2019,2020JCAP...03..044N}\footnote{https://github.com/LSSTDESC/NaMaster}.
The pseudo-$C_\ell$ estimator, as well as many other cosmic shear statistics~\citep[e.g.,][]{2018MNRAS.478.4277S}, typically employs inverse-variance weighting based on the intrinsic shape noise of source galaxies.
However, this weighing scheme is optimal only on small angular scales, where intrinsic shape noise dominates. 
It is therefore not necessarily optimal on large angular scales, where sample variance dominates. 
To fully exploit the statistical power of ongoing and upcoming galaxy surveys, such as the Vera C. Rubin Observatory's Legacy Survey of Space and Time (LSST)\footnote{https://www.lsst.org} and the Nancy Grace Roman Space Telescope\footnote{https://science.nasa.gov/mission/roman-space-telescope/}, it is important to develop a statistically optimal estimator applicable across all angular scales.

The purpose of this paper is therefore to develop an optimal estimator for the cosmic shear power spectrum that is statistically optimal across all angular scales by using a covariance matrix incorporating both sample variance and intrinsic shape noise~\citep[see also Ref.][for a similar study focused on large angular scales]{maraio_testing_2023}.
We derive a quadratic estimator for the cosmic shear power spectrum by maximizing the Gaussian likelihood of galaxy shape field~\citep[see Ref.][for the CMB study]{oh_efficient_1999}. 
By adopting the flat-sky approximation and extending pioneering methods developed for CMB \citep{smith_algorithms_2011} and galaxy-clustering analyses~\citep{philcox_cosmology_2021,philcox_cosmology_2021-1,philcox_optimal_2023,philcox_polybin3d_2025}, we render the inverse-covariance matrix operations computationally tractable.
In doing so, we ensure that the estimator suppresses the leakage of $E$-mode power into the $B$-mode power spectrum more effectively than the pseudo-$C_\ell$ method.
We then validate the method and assess its performance using simulated galaxy shape fields generated from both Gaussian shear fields and from cosmological ray-tracing simulations of a $\Lambda$CDM model.

This paper is organized as follows.
In Section~\ref{sec:method}, we derive an optimal quadratic estimator for the $E$- and $B$-mode cosmic shear power spectra by maximizing the Gaussian likelihood function of galaxy shape field.
In Section~\ref{sec:validation}, we validate the method and evaluate its performance using Gaussian simulations of galaxy shape fields.
In Section~\ref{sec:raytracing}, we apply the method to more realistic galaxy shape fields generated by ray-tracing simulations of a $\Lambda$CDM cosmology and further evaluate its performance.
Finally, Section~\ref{sec:conclusion} presents our discussion and conclusions.

Throughout this paper, we adopt the following condensed notations for simplicity:
\begin{align}
\int_{\B{\ell}}\equiv\int\!\frac{\mathrm{d}^2\B{\ell}}{(2\pi)^2},
\hspace{1em}
\int_{\boldsymbol{\theta}}\equiv \int\!\mathrm{d}^2\boldsymbol{\theta}
\end{align}

Here, we define the forward and backward Fourier transforms as
\begin{align}
    \mathcal{F}[a](\B{\ell}) = \int_\B{\theta}\ e^{-i\B{\ell}\cdot\B{\theta}}a(\B{\theta}) \ , \ \mathcal{F}^{-1}[b](\B{\theta}) = \int_{\B{\ell}}\ e^{i\B{\ell}\cdot\B{\theta}}\ b(\B{\ell}), \label{eq:fourier}
\end{align}
for given real-space field $a(\B{\theta})$ and  Fourier-space field $b(\B{\ell})$.

\section{Methodology for Measuring the Cosmic Shear Power Spectrum}
\label{sec:method}

In this section, we describe an estimator for cosmic shear power spectrum under the flat-sky approximation. 
Our derivation is based on the previous works~\citep{bond_estimating_1998, oh_efficient_1999,smith_algorithms_2011,
philcox_cosmology_2021, terawaki_quadratic_2025}.

\subsection{Flat-sky formalism}
\label{subsec:flatsky}
Throughout this paper, we adopt the flat-sky approximation.
Under this approximation, we can expand the shear fields (or more generally galaxy shape fields) $\gamma_\pm\equiv \gamma_1\pm i\gamma_2$ using the Fourier transforms as follows:
\begin{align}
    \gamma_\pm(\B{\theta})=\int_\B{\ell}e^{i\B{\ell}\cdot\B{\theta}}e^{\pm2i\phi_\B{\ell}}[E(\B{\ell})\pm iB(\B{\ell})]
, \label{eq:shear}
\end{align}
where $E/B$ mode denotes the parity even/odd components of the field and $\phi_\B{\ell}$ denotes 
the power angle for the wavenumber vector $\B{\ell}$, defined as $\B{\ell}=\ell(\cos\phi_\B{\ell},\sin\phi_\B{\ell})$.

In an actual survey, the observed field is affected by a survey window, as given by $\gamma^W_\pm(\B{\theta})=W(\B{\theta})\gamma_\pm(\B{\theta})$, where $W(\B{\theta})$ is the survey window function that accounts for the survey geometry and star masks; $W(\B{\theta})=1$ if $\B{\theta}$ is inside the survey region, and $W(\B{\theta})=0$ otherwise. 
We do not consider a weighting function for the data vector here, but we will discuss in a later section. 
Then, the covariance matrix of the observed shear field is given as
\begin{align}
    &C^{ij}(\B{\theta_1},\B{\theta_2})\equiv
     \ev{\gamma^W_i(\B{\theta_1})\gamma^W_j(\B{\theta_2})}\nonumber\\
    &\qquad=W(\B{\theta_1})W(\B{\theta_2})\int_{\B{\ell}}e^{i\B{\ell}\cdot(\B{\theta_1}-\B{\theta_2})}\sum_{s}f^{ij}_s(\phi_\B{\ell})\mathcal{C}^s(\ell)\nonumber\\
    &\qquad\qquad\qquad\qquad\qquad\qquad+\mathcal{N}^{ij}(\B{\theta_1},\B{\theta_2}), \label{eq:cov}
\end{align}
where the indices $i,j \in \{+,-\}$ denote the two components of the shear fields, and $s\in\{\rm{EE},\rm{EB},\rm{BB}\}$ denote the corresponding mode power spectra.
$\mathcal{C}^{s}(\ell)$ represents the angular power spectra and $\mathcal{N}^{ij}$ is a shape noise contribution. 
The phase function $f^{ij}_s(\phi_\B{\ell})$, which satisfies the Hermitian condition $f^{ij}_s(\phi_\B{\ell}) = f^{ji,*}_s(\phi_\B{\ell})$, is defined as
\begin{align}
    &f^{\pm\pm}_{\rm{EE}}(\phi_\B{\ell})=f^{\pm\pm}_{\rm{BB}}(\phi_\B{\ell})=1  \ , \nonumber \\ 
    &f^{\pm\pm}_{\rm{EB}}(\phi_\B{\ell})=0 \ ,
    \nonumber\\
    &f^{+-}_{\rm{EE}}(\phi_\B{\ell})=-f^{+-}_{\rm{BB}}(\phi_\B{\ell})=e^{4i\phi_\B{\ell}} \ , \nonumber\\
    &f^{+-}_{\rm{EB}}(\phi_\B{\ell})=2ie^{4i\phi_\B{\ell}} \ , 
    \label{eq:phase}
\end{align}
where the signs are in the same order.
Introducing the band power parameters $\B{p}=\{p_{s;\alpha}\}$, we can express the power spectrum as a discrete sum:
$\mathcal{C}^s(\ell)=\sum_{\alpha} p_{s;\alpha} \Theta_\alpha(|\B{\ell}|)$, where $\Theta_\alpha(|\B{\ell}|)$ is the Heaviside step function; 
$\Theta_\alpha(|\B{\ell}|)=1$ if $\B{\ell}$ is inside the $\alpha$-th $\ell_\alpha$ bin and $\Theta_\alpha(|\B{\ell}|)=0$ otherwise.
Here we assume that the power spectrum varies slowly enough within each $\ell_\alpha$ bin.
In this setup, we can compute the partial derivative of the covariance matrix with respect to the $\alpha$-th and mode-$s$ band power parameter as
\begin{align}
    &C^{ij}_{,s;\alpha}(\B{\theta_1},\B{\theta_2})\equiv
    \pdv{C^{ij}(\B{\theta_1},\B{\theta_2})}{p_{s;\alpha}}
    \nonumber\\
    &\qquad =W(\B{\theta_1})W(\B{\theta_2})\int_{\B{\ell}}e^{i\B{\ell}\cdot(\B{\theta_1}-\B{\theta_2})}f^{ij}_s(\phi_\B{\ell})\Theta_\alpha(\ell) . \label{eq:cov_derivative}
\end{align}
Note that the derivative of the covariance matrix is independent of the underlying cosmological model.

\subsection{Maximum likelihood estimator}
\label{subsec:ME}
In this section, we present the maximum likelihood estimator (MLE) for cosmic shear power spectrum, which is the primary focus of this paper.

\subsubsection{Likelihood formalism}
\label{subsubsec:likelihood}
Let us consider shear fields defined on pixels and denote the total number of pixels as $N_{\mr{pix}}$. 
For the weak lensing case, we define the $2N_{\rm{pix}}$-dimensional shear data vector as $\B{d}\equiv (\B{d}_+,\B{d}_-)$, whose components are defined as
\begin{align}
    \B{d}_\pm=\left\{\gamma^W_\pm(\B{\theta_1}),\cdots,\gamma^W_\pm(\B{\theta}_{N_{\mr{pix}}})\right\}. \label{eq:datavector}
\end{align}
Then, since the shear has two components,
the covariance matrix $C= \ev{\B{d}\B{d}^\dagger}$ consists of $2\times 2$ block matrices, such as $\langle\B{\gamma}_+\B{\gamma}_+^\dagger\rangle$,
$\langle\B{\gamma}_+\B{\gamma}_-^\dagger\rangle$, 
$\langle\B{\gamma}_-\B{\gamma}_+^\dagger\rangle$, 
and 
$\langle \B{\gamma}_-\B{\gamma}_-^\dagger\rangle$.
Each block matrix has dimensions of $N_{\mr{pix}}\times N_{\mr{pix}}$. 
We typically adopt $N_{\rm pix}\sim O(10^{6})$, so the covariance matrix has a very large dimension.

A maximum likelihood estimator requires the covariance matrix.
However, because the survey window introduces zero entries, the covariance matrix $C$ is generally singular and non-invertible.
Therefore, we define subspace data vector $\tilde{\B{d}}$ and matrices $\tilde{C},\tilde{\mathcal{N}},\cdots$ by retaining only the components where $W(\boldsymbol{\theta}) \neq 0$, 
thereby ensuring that the inverse matrix $\tilde{C}^{-1}$ exists.
Note that $\tilde{C}$ has dimension $N_{\rm sub}\times N_{\rm sub}$, where $N_{\rm sub}$ is the number of pixels with $W(\B{\theta})\neq 0$ in the shear field map; therefore $N_{\rm sub}\le N_{\rm pix}$.
Then, we can safely define the likelihood function assuming a Gaussian likelihood function as
\begin{align}
\mathcal{L}(\tilde{\B{d}}|\B{p})=\B{\tilde{d}}^\dagger \tilde{C}_{\mr{true}}^{-1}(\B{p})\B{\tilde{d}}+\ln (\mr{det}[\tilde{C}_{\mr{true}}(\B{p})])+\mr{const.} ,\label{eq:likelihood}
\end{align}
where $\tilde{C}_{\mr{true}}$ is the true covariance in the subspace containing non-zero elements.\footnote{Note that our formalism differs slightly from standard approaches, such as those in \cite{philcox_cosmology_2021, philcox_polybin3d_2025}, which employ the full covariance matrix in the likelihood function.}
Taylor-expanding around fiducial band power in each multipole bin, $\B{p}_{\rm{fid}}$, and finding a maximum likelihood solution, we obtain the maximum likelihood estimator (see \cite{philcox_cosmology_2021,terawaki_quadratic_2025} for details):
\begin{align}
    \hat{p}_{s;\alpha}&=\sum_\beta\sum_t (F^{-1})_{st;\alpha\beta}(\hat{q}_{t;\beta}-\hat{n}_{t;\beta}) \ ,
    \label{eq:MLE}
\end{align}
with
\begin{align}
        F_{st;\alpha\beta}&=\frac{1}{2}\mr{Tr}\sqbra{\tilde{C}_{\mr{fid}}^{-1}\tilde{C}_{,s;\alpha}\tilde{C}_{\mr{fid}}^{-1}\tilde{C}_{,t;\beta}} \ , \nonumber\\
    \hat{q}_{s;\alpha}&=\frac{1}{2}\mr{Tr}\sqbra{\tilde{C}_{\mr{fid}}^{-1}\tilde{C}_{,s;\alpha}\tilde{C}_{\mr{fid}}^{-1}\B{\tilde{d}}\B{\tilde{d}}^\dagger} \ ,\nonumber\\
    &=\frac{1}{2}\parbra{\tilde{C}_{\mr{fid}}^{-1}\B{\tilde{d}}}^\dagger \tilde{C}_{,s;\alpha}\parbra{\tilde{C}_{\mr{fid}}^{-1}\B{\tilde{d}}} \ , \nonumber\\
    \hat{n}_{s;\alpha}&=\frac{1}{2}\mr{Tr}\sqbra{\tilde{C}_{\mr{fid}}^{-1}\tilde{C}_{,s;\alpha}\tilde{C}_{\mr{fid}}^{-1}\tilde{\mathcal{N}}} .
    \label{eq:components}
\end{align}
Here $\hat{n}$ is the shape noise contribution to the data term $\hat{q}$, and the matrix $F$ is the Fisher matrix, defined as the second derivatives of the likelihood with respect to the band powers. 
We do not a priori know the covariance $C$ of the true power spectrum, which we aim to estimate; 
therefore, in the above estimator, we use the inverse of the covariance matrix for a fiducial cosmology, denoted as $\tilde{C}^{-1}_{\rm fid}$.
In practice, the estimator can be iterated until convergence is reached (until $\tilde{C}^{-1}_{\rm fid}$ is sufficiently close to the true one).
Note that, for each $s\in \{{\rm EE, EB, BB}\}$ each of $\hat{p}_s, \hat{q}_s$, and $\hat{n}_s$ is a $N_{\rm bin}$-dimension vector corresponding to the band power labeled by $s$, $\B{F}_{st}$ is a $N_{\rm bin}\times N_{\rm bin}$-dimension matrix for each $(s,t)\in \{{\rm EE, EB, BB}\}$, where $N_{\rm{bin}}$ denotes the number of multipole bins.

Eq.~(\ref{eq:MLE}) represents the ML estimator for the cosmic shear power spectrum, which is the main focus of this paper. 
We will numerically implement this estimator to demonstrate that it is indeed an optimal estimator of the underlying power spectrum, compared with the pseudo-$C_\ell$ estimator commonly used in the literature.

We can easily check that Eq.~(\ref{eq:MLE}) is an unbiased estimator of the true band power in the ensemble average sense:
\begin{align}
\avrg{\hat{p}_{s;\alpha}}&=\sum_{\beta,t}(F^{-1})_{st;\alpha\beta}
\avrg{\hat{q}_{t;\beta}-\hat{n}_{t;\beta}}\nonumber\\
&= \sum_{\beta, t}(F^{-1})_{st;\alpha\beta} \frac{1}{2}
{\rm Tr}\left[\tilde{C}^{-1}_{\rm fid}
\tilde{C}_{,t;\beta}\tilde{C}^{-1}_{\rm fid}
\avrg{\tilde{\B{d}}\tilde{\B{d}}^\dagger-\tilde{{\cal N}}}
\right]\nonumber\\
&= \sum_{\beta, t}(F^{-1})_{st;\alpha\beta} \frac{1}{2}
{\rm Tr}\left[\tilde{C}^{-1}_{\rm fid}
\tilde{C}_{,t;\beta}\tilde{C}^{-1}_{\rm fid}
\sum_{s',\mu}\tilde{C}_{,s';\mu}p^{\rm true}_{s';\mu}\right]\nonumber\\
&=\sum_{\beta,t}\sum_{\mu, s'}
(F^{-1})_{st;\alpha\beta} F_{ts';\beta\mu}p^{\rm true}_{s';\mu}=p^{\rm true}_{s;\alpha}\ ,
\label{eq:unbiased_proof}
\end{align}
where we used the definition of the Fisher matrix, given by 
Eq.~(\ref{eq:components}), to rewrite the third line on the r.h.s.
We again stress that this estimator provides an estimate of the window-free band power.
The vector $\hat{q}-\hat{n}$ corresponds to the window-convolved power spectrum, including the effects of the mode-coupling and $E/B$-mode mixing due to the survey window.
Therefore, as can be seen from the above equation, the inverse of Fisher matrix, $F^{-1}$, correct for the window effects.

The covariance in Eq.~(\ref{eq:components}) has two contributions, given by $\tilde{C}_{\rm fid}=\tilde{S}_{\rm fid}+\tilde{N}$, where $\tilde{S}_{\rm fid}$ and $\tilde{N}$ are the sample variance and shape noise contributions, respectively.
We note that $\tilde{C}^{-1}_{\rm fid}$ in $(\tilde{C}^{-1}_{\rm fid}\tilde{\B{d}})$ of Eq.~(\ref{eq:components}) can be interpreted as a weighting of the data vector $\tilde{\B{d}}$.
Therefore, since $\tilde{C}_{\rm fid}\simeq \tilde{S}_{\rm fid}$ and $\tilde{C}_{\rm fid}=\tilde{{\cal N}}$ in the sample variance and shape noise dominated regimes, respectively, the estimator is optimal in both regimes, as we will demonstrate quantitatively below. 

\subsubsection{Practical implementation}
\label{subsubsec:implement}

A naive implementation of Eq.~(\ref{eq:MLE}) is computationally expensive due to the large number of pixels (typically $N_{\rm pix}\sim 10^6$); e.g., a direct computation of the Fisher matrix $\B{F}$ in Eq.~(\ref{eq:components}) requires $O(N_{\rm pix}^3)$ matrix operations since $N_{\rm sub}\sim N_{\rm pix}$, and this is therefore computationally intractable.
Instead, we use the fast Fourier transform (FFT) method~\citep{philcox_cosmology_2021,philcox_cosmology_2021-1} to efficiently perform these computations.
However, FFTs operate on rectangular grids, requiring zero-padding for any pixels without valid data. 
Consequently, the calculations for Eq.~(\ref{eq:MLE}) must be formulated in the full-space domain, rather than being restricted to the unmasked subspace. As an example, the data term is rewritten as:
\begin{align}
    \hat{q}_{s;\alpha}
    &=\frac{1}{2}\parbra{C^{-1}_{\mr{fid}}\B{d}}^\dagger C_{,s;\alpha}\parbra{C^{-1}_{\mr{fid}}\B{d}}\label{eq:QE_full}
\end{align}
Here, $C_{\rm fid}^{-1}(\boldsymbol{\theta}, \boldsymbol{\theta}')$ is defined to coincide with the subspace weight $\tilde{C}_{\rm fid}^{-1}$ for pixels within the survey footprint, while being set to zero elsewhere.
This definition of the inverse matrix satisfies a modified invertibility condition, expressed as $C_{\rm fid}^{-1}({\B{\theta}_1,\B{\theta}_2})C_{\rm fid}({\B{\theta}_2,\B{\theta}_3})=W(\B{\theta}_1)\delta^K_{\B{\theta}_1,\B{\theta}_3}$.
Then, this full-space estimator actually becomes equivalent to the subspace estimators in Eqs.~(\ref{eq:MLE}) and (\ref{eq:components}).

As shown in Eq.~(\ref{eq:cov_derivative}), the derivative of the covariance $C_{,s;\alpha}$ is expressed as the operation of window matrices.
This implies that the only required quantity is the product of the window matrix and the weighted data, which we define as the windowed-weighted data: 
$h_\pm(\B{\theta}) \equiv W(\B{\theta})(C^{-1}_{\rm fid}\B{d})_\pm(\B{\theta})$.
However, implementing this optimal weighting scheme is generally infeasible due to the need to invert a prohibitively large covariance matrix.
To overcome this difficulty, we note that the matrix $C_{\rm fid}^{-1}$ appears only in combination with a vector (and the noise term $\hat{n}$ and Fisher matrix $F$ can be similarly reformulated to satisfy this condition, as we will show later).
We therefore employ the conjugate gradient (CGM) method to numerically solve the linear equation $A\B{y} = \B{x}$ for the vector $\B{y}$, following Ref.~\cite{smith_algorithms_2011}\citep[also see][]{philcox_cosmology_2021,terawaki_quadratic_2025}.
Since the CGM algorithm relies solely on repeated matrix-vector multiplications, solving the linear system $\B{d}=C_{\rm fid}\B{m}$ in the full-space is numerically equivalent to solving $\tilde{\B{d}}=\tilde{C}_{\rm fid}\tilde{\B{m}}$ in the subspace. 
This equivalence -- further supported by the fact that we only require the windowed vector $h_\pm(\B{\theta}) = W(\B{\theta})\B{m}_\pm(\B{\theta})$ -- justifies our use of the full-space framework throughout the analysis.

To compute $\hat{\B{q}},\hat{\B{n}}, \B{F}$, we follow the same procedure written in Ref.~\cite{terawaki_quadratic_2025}. 
In this paper, we briefly summarize the implementation using FFTs.
We can use the following FFT method to compute the data term $\hat{q}$:
\begin{align}
    \hat{q}_{s;\alpha}=\frac{1}{2}\sum_{ij}\int_\B{\ell}\Theta_\alpha(\ell)f^{ij}_s(\phi_\B{\ell})
    {\cal F}^*[h_i](\B{\ell}){\cal F}[h_j](\B{\ell}). \label{eq:first_imple}
\end{align}
Here, the FFT computation, denoted by ${\cal F}$, involves $O(N_{\rm pix}\ln N_{\rm pix})$ matrix operation, and then the integral in the $\B{\ell}$ space involves $O(N_{\rm FFT}^2)$, while $N_{\rm FFT}$ is the number of FFT grids. 
Thus, this computation is feasible.
However, the naive expression for the Fisher matrix does not explicitly involve such vector forms, and is therefore not straightforward to compute.
In this paper, as shown in Refs.~\cite{oh_efficient_1999,smith_algorithms_2011,philcox_cosmology_2021,philcox_cosmology_2021-1}, we use the Monte Carlo method by introducing ``ancillary'' Gaussian fields, denoted as $\{\B{a}\}$.
These fields satisfy the relation $\ev{\B{a}\B{a}^\dagger}=C_{\rm{fid}}$, where the notation $\langle\hspace{0.5em}\rangle$ denotes the ensemble average, which is practically computed using a sufficient number of realizations of the simulated $\{\B{a}\}$ fields.
Here, we define the windowed-weighted ancillary fields as $\B{u}_\pm(\B{\theta})\equiv W(\B{\theta})(C_{\rm fid}^{-1}\B{a})_\pm(\B{\theta})$.
Then, we can evaluate the Fisher matrices numerically in an ensemble average sense:
\begin{align}
    F_{st;\alpha\beta}&=\frac{1}{2}\ev{g^\dagger_{s;\alpha} C^{-1}_{\mr{fid}}g_{t;\beta}}, \label{eq:fisher_imple}
\end{align}
where we define the real vector $\B{g}_{s;\alpha}$ as follows:
\begin{align}
    \B{g}_{i,s;\alpha}(\B{\theta})&\equiv W(\B{\theta})\sum_j \mathcal{F}^{-1}\Big{[}\Theta_\alpha f^{ij}_s{\cal F}[u_j]\Big](\B{\theta}).
    \label{eq:gfield}
\end{align}
For the noise term $\hat{n}$, we can rewrite the noise covariance matrix using another ancillary vectors $\B{n}$ as $\mathcal{N}=\ev{\B{n}\B{n}^\dagger}$. 
Then, $\hat{n}$ can be computed in a completely the same manner as Eq.~(\ref{eq:first_imple}).

\subsection{Optimality} 
\label{sec:optimality}

In this subsection, we comment on the fact that the estimator given in Eq.~(\ref{eq:MLE}) is indeed optimal.
To do so, we first note that $\tilde{C}^{-1}_{\rm fid}$ in $(\tilde{C}^{-1}_{\rm fid}\tilde{\B{d}})$ of Eq.~(\ref{eq:components}) can be interpreted as a weighting of the data vector $\tilde{\B{d}}$.
Therefore, we can write down a general quadratic estimator which can be obtained by replacing all $\tilde{C}_{\rm{fid}}^{-1}$ in the equations of Eq.~(\ref{eq:components}) with an arbitrary matrix~$\tilde{H}^{-1}$. 
In this case, the data term of the general quadratic estimator is given as
\begin{align}
    \hat{q}_{s;\alpha}=\frac{1}{2}\parbra{\tilde{H}^{-1}\B{\tilde{d}}}^\dagger \tilde{C}_{,s;\alpha}\parbra{\tilde{H}^{-1}\B{\tilde{d}}}
    \label{eq:QE}
\end{align}
One can readily check that the quadratic estimator using $H^{-1}$
yields an unbiased estimate of the true band power.

The statistical errors of the estimated band powers are described by the covariance matrix, which is given as
\begin{align}
    &\mr{Cov}(\hat{p}_{s;\alpha},\hat{p}_{t;\beta})\nonumber\\
    &\hspace{1em}=\sum_{\gamma,\delta}\sum_{u,v}(F^{-1})_{su;\alpha\gamma}(F^{-1})_{tv;\beta\delta}\mr{Cov}(\hat{q}_{u;\gamma},\hat{q}_{v;\delta})\ .
    \label{eq:cov_p_def}
\end{align}
Assuming that the data follow a Gaussian distribution, the covariance of $\hat{q}$ in the above equation is given by 
\begin{align}
    &\mr{Cov}(\hat{q}_{u;\gamma},\hat{q}_{v;\delta})\nonumber\\
    &\hspace{1em}=\frac{1}{2}\mr{Tr}\sqbra{(\tilde{H}^{-1}\tilde{C}_{\mr{true}}\tilde{H}^{-\dagger}) \tilde{C}_{,u;\gamma}(\tilde{H}^{-1}\tilde{C}_{\mr{true}}\tilde{H}^{-\dagger})\tilde{C}_{,v;\delta}}
    \nonumber\\
    &\hspace{1em}\equiv\mathbb{F}[\tilde{H}^{-1}\tilde{C}_{\mr{true}}\tilde{H}^{-\dagger}]
     ,\label{eq:cov_q}
\end{align}
where we have introduced the notation $\mathbb{F}[M]$ to denote that the Fisher matrix is given by replacing $C_{\rm fid}^{-1}$ in Eq.~(\ref{eq:components}) with a matrix $M$.
From the Cram\'er-Rao theorem, we can find that the following inequality for the covariance for $\hat{\B{p}}$ holds
\begin{align}
    \mathbb{F}^{-1}[\tilde{H}^{-1}]\mathbb{F}[\tilde{H}^{-1}\tilde{C}_{\mr{true}}\tilde{H}^{-\dagger}]\mathbb{F}^{-1}[\tilde{H}^{-1}] \geq \mathbb{F}^{-1}[\tilde{C}_{\mr{true}}^{-1}] . \label{eq:optimal_condition}
\end{align}
From Eqs.~(\ref{eq:cov_p_def}) and (\ref{eq:optimal_condition}), we can find
\begin{align}
{\rm Cov}(\hat{p}_{s;\alpha},\hat{p}_{t;\beta})\ge \mathbb{F}^{-1}[\tilde{C}_{\mr{true}}^{-1}]\ . 
\end{align}
Thus, {\it only} by setting $\tilde{H}^{-1}=\tilde{C}_{\rm true}^{-1}$, we can have a minimum-variance estimator for the band power.

\subsection{A suboptimal estimator: pseudo-$C_\ell$}
\label{sec:pseudo-Cell}

The pseudo-$C_\ell$ estimator \citep{hikage_shear_2011,alonso_unified_2019} is a method commonly used in cosmic shear analyses~\cite{doux_dark_2022,dalal_hyper_2023}.
The pseudo-$C_\ell$ estimator can be obtained by adopting a diagonal matrix for the weighting matrix $\tilde{H}^{-1}$ in Eqs.~(\ref{eq:MLE}) and (\ref{eq:components}), as shown in our previous work \cite{terawaki_quadratic_2025}. 
In particular, the following weighting is often adopted:
\begin{align}
\tilde{H}^{-1}\rightarrow
\tilde{\mathcal{N}}^{-1}(\B{\theta},\B{\theta}')&=\frac{1}{\sigma^2_{\mr{int},p}(\B{\theta})+\sigma^2_{\mr{meas},p}(\B{\theta})}{\delta}^K_{\B{\theta},\B{\theta}'}
.\label{eq:shape_weight}
\end{align}
where ${\delta}^K_{\B{\theta},\B{\theta}'}$ is the Kronecker delta function; $\delta^K_{\B{\theta},\B{\theta}'}=1$ if $\B{\theta}=\B{\theta}'$, and $\delta^K_{\B{\theta},\B{\theta}'}=0$ otherwise.
Here, $\sigma_{\mr{int},p}$ and $\sigma_{\mr{meas},p}$ are defined as the rms intrinsic ellipticity and the measurement error in each pixel, obtained by averaging over galaxies in each pixel.
The diagonal form of $\tilde{H}^{-1}$ enables fast computation of the ingredients in the estimator (Eqs.\ref{eq:MLE} and \ref{eq:components}), such as the Fisher matrix, without the need to use CGM. 
We also note that, as discussed below Eq.~(\ref{eq:unbiased_proof}), the above choice corresponds to the limit of $\tilde{C}^{-1}_{\rm true}\rightarrow \tilde{{\cal N}}^{-1}$ in the shape noise dominated limit.
Therefore, the pseudo-$C_\ell$ estimator using this $\tilde{H}^{-1}=\tilde{\cal N}^{-1}$ is optimal only in the shape noise dominated regime, but is sub-optimal in the sample variance limited regime.
As a benchmark test of the maximum likelihood estimator (MLE), we also consider the pseudo-$C_\ell$ estimator and compare its results with those from MLE.

\section{Validation with Gaussian random fields}
\label{sec:validation}

In this section, we apply our method to Gaussian random fields to validate its performance.
Furthermore, since this setup represents the optimal limit, we quantify the maximum potential gain achievable with the optimal weighting.

\subsection{Setups for the validation}
\label{subsec:setups}
\begin{figure}[h]
\centering 
\includegraphics[width=0.48\textwidth]{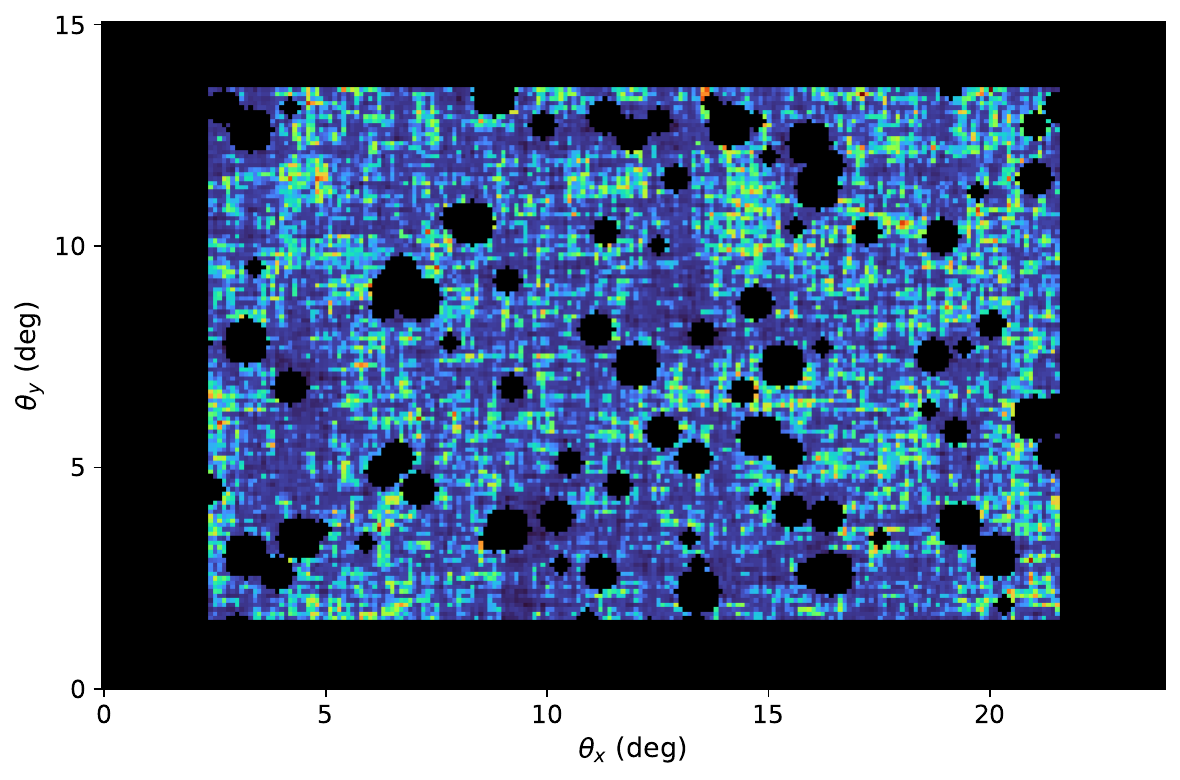}
\caption{
An example of a simulated Gaussian shear field generated under the flat-sky approximation.
The color scale shows one of the spin-2 shear components, $\gamma_1$.
The simulated field is defined over a rectangular region of $15 \times 24 \, \text{deg}^2$, discretized into $150 \times 240$ pixels with a pixel resolution of $6 \, \text{arcmin}$.
We first mask the outermost 10\% of the map along both the $\theta_x$- and $\theta_y$-directions and apply zero padding to these regions, thereby breaking the periodic boundary conditions.
We then add circular masks with radii of up to 30~arcmin that mimic star masks, and these masks cover approximately 15\% of the data region.
}
\label{fig:window}
\end{figure}
To evaluate the performance of our optimal estimator, we use Gaussian random fields in the flat-sky approximation.
Since the optimality condition is derived under the assumption of Gaussianity, this setup provides an ideal benchmark for quantifying the maximum achievable improvement in statistical precision.
We first construct simulated Gaussian shear maps, each covering a rectangular field of $24^\circ\times 15^\circ =360 \, \text{deg}^2$, discretized into $240\times 150$ pixels with a pixel size of $6 \, \text{arcmin}$.
In this setup, the fundamental mode for FFT, corresponding to the largest scale of the field, is defined by $\ell_{\text{fund}} \approx 2\pi/L$, yielding $\ell_{\text{min}, x} \simeq 24$ and $\ell_{\text{min}, y} \simeq 15$. 
We generate 1,000 realizations of the Gaussian shear fields $\gamma_1$ and $\gamma_2$ from the input $E$-mode power spectrum $C^{\rm EE}_\ell$.
This input $E$-mode spectrum is computed using \texttt{ccl} for a single source redshift slice at $z=1$, assuming a fiducial $\Lambda$CDM model with the following parameters: $\Omega_{\rm c}=0.233, \Omega_{\rm b}=0.046$, $\Omega_\Lambda=0.721, h=0.7, n_s=0.97$, and $\sigma_8=0.82$.
Note that the input $B$-mode spectra, $C^{\rm{EB}}_\ell$ and $C^{\rm{BB}}_\ell$, are set to zero.

To mimic realistic survey data, we introduce masked regions into each simulated map.
We first mask the outermost 10\% of the map along both the $\theta_x$- and $\theta_y$-directions and apply zero padding to these regions, as shown in Fig.~\ref{fig:window}. 
This operation breaks the periodic boundary conditions of the Gaussian shear maps. 
As a result, the effective area containing shear data is reduced to
$19.2^\circ\times 12^\circ\simeq 230.4~{\rm deg}^2$. 
We then add small-scale masks mimicking bright star masks.
We randomly distribute circular masks with radii up to 30~arcmin. 
The total masked areas is set to approximately 15\% of the data region, a value typical of actual weak lensing surveys \citep[e.g.,][]{li_three-year_2022}. 
Note that we employ a binary window function $W(\B{\theta}) \in \{0, 1\}$ on each pixel without applying any apodization.

We further include shape noise in each simulated map, arising from the
intrinsic ellipticities $\epsilon_{\mr{int},i}$ and measurement errors $\epsilon_{\mr{meas},i}$ of individual galaxies.
This shape noise introduces a white noise component into the measured power spectrum.
The shape noise variance of individual galaxies is determined by various observational factors, such as the detection signal-to-noise ratio and the galaxy size relative to the point spread function (PSF)~\citep[see, e.g.,][]{li_three-year_2022}.
Furthermore, for pixel-based analyses, the effective noise variance in each pixel scales inversely with the number of galaxies within the pixel.
In this analysis, we construct inhomogeneous variance maps to make a realistic noise distribution.
We define the $1\sigma$ uncertainty for $\epsilon_{\mr{int},i}$ and $\epsilon_{\mr{meas},i}$ in each pixel at position $\B{\theta}$ as follows:
\begin{align}
\sigma_{c,p}(\B{\theta})=\frac{\sigma_{c,g}(\B{\theta})}{\sqrt{N_{\mr{gal}}(\B{\theta})}}, 
\label{eq:noise per pixel}
\end{align}
where $c\in \{\rm{int, meas}\}$, $\sigma_{c,g}$ denotes the $1\sigma$ uncertainty per galaxy (Eq.~\ref{eq:shape_weight}), 
and $N_{\rm{gal}}$ is the number of galaxies in the pixel.
Following the configurations for HSC galaxies in \citet[Fig.~9]{li_three-year_2022}, 
we generate spatially varying maps of $\sigma_{\mathrm{int},g}$ and $\sigma_{\mathrm{meas},g}$ by independently drawing the value in each pixel from uniform distributions over the ranges $[0.37, 0.43]$ and $[0.05, 0.3]$, respectively.
The galaxy counts in each pixel, $N_{\rm{gal}}(\B{\theta}_{\rm{pix}})$, is drawn from a Poisson distribution with mean $\bar{n}_s\Omega_{\rm{pix}}$, where $\bar{n}_s$ and $\Omega_{\rm{pix}}$ denote the mean number density of source galaxies and the pixel area, respectively.
To investigate how the gain of optimal weighting depends on the shape noise level, we adopt four galaxy number densities, $\bar{n}_s = 8, 20, 30$ and $40\ \rm{arcmin}^{-2}$, representing the typical values of the DES~\citep{collaboration_dark_2026,giannini_dark_2026}, HSC~\citep{hikage_cosmology_2019,dalal_hyper_2023}, LSST~\citep{fang_cosmology_2021}, and Roman surveys~\citep{cao_fisher_2026}.
Then, for each map, the noise components $\epsilon_{c,i}(\B{\theta}_{\rm{pix}})$ are generated by drawing Gaussian random variates from $\mathcal{N}(0, \sigma^2_{c,p}(\B{\theta}_{\rm{pix}}))$. 
These components are added to the shear signal $\gamma_i$, yielding the simulated data: 
$d_i = \gamma_i + \epsilon_{\mr{int},i}+ \epsilon_{\mr{meas},i}$.
We also use the shape-noise-only field, $n_i=\epsilon_{\mr{int},i}+ \epsilon_{\mr{meas},i}$, to estimate the noise power spectrum.

Following the discussion in Sections~\ref{sec:optimality} and \ref{sec:pseudo-Cell}, we compare the performance of the optimal weighting scheme with the suboptimal weighting scheme.
The suboptimal weight, $\mathcal{N}^{-1}$, is computed by substituting Eq.~(\ref{eq:noise per pixel}) into Eq.~(\ref{eq:shape_weight}). 
For the optimal weight, defined as $C_{\rm{fid}}^{-1} = (S_{\rm{fid}} + \mathcal{N})^{-1}$, we adopt the true angular power spectrum as the fiducial signal $S_{\rm{fid}}$, namely the same spectrum used to generate the underlying shear fields.
This setup therefore represents an idealized scenario that allows us to quantify the maximum achievable performance gain from optimal weighting.

\subsection{Performance of window-free power spectrum estimation}
\label{subsec:power measure}
\begin{figure}[h]
\centering 
\includegraphics[width=0.45\textwidth]{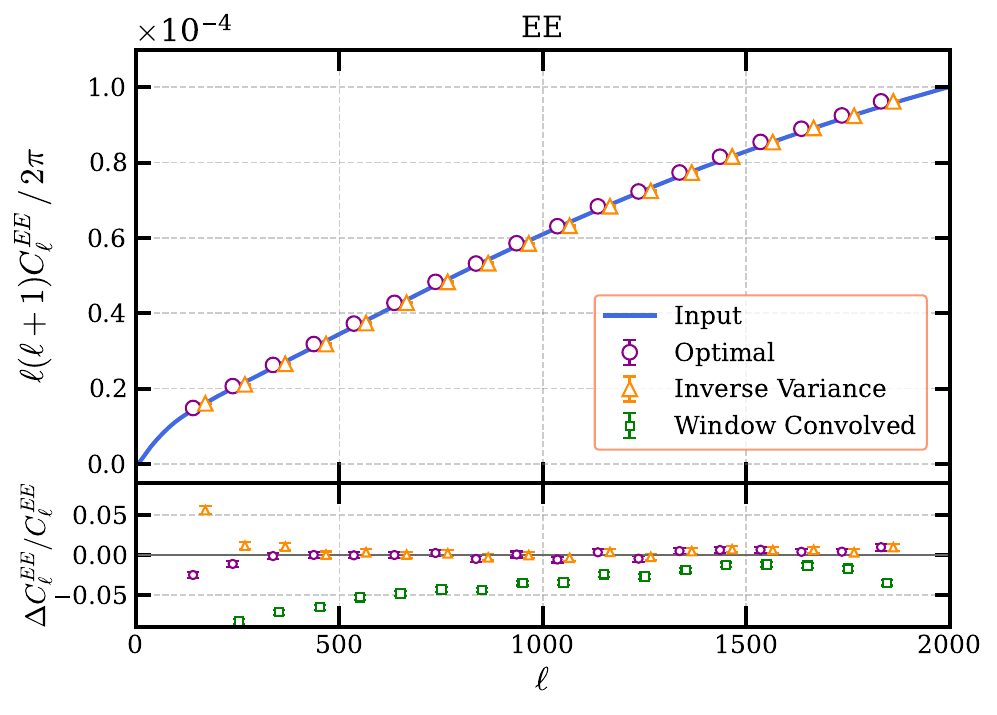}
\caption{
{\it Upper panel}: Angular power spectrum $C^{\mathrm{EE}}_\ell$ measured by applying the quadratic estimator to Gaussian shear fields over a 360~deg$^2$ area as shown in Fig.~\ref{fig:window}. 
We include the HSC-like shape noise contribution to the shear fields (see text for details).
The purple circles and orange triangles represent the results obtained using the optimal weighting (this work)
and the inverse variance weighting based on shape noise, respectively. 
Each data point indicates the mean band power averaged over 1,000 realizations, with error bars denoting the $1\sigma$ uncertainty on the mean.
Although we use the same wavenumber bins for both weighting schemes, we slightly shift the data points for easier comparison.
For comparison, the blue line show the input power spectrum used to create the Gaussian shear fields.
{\it{Lower panel}}: The fractional difference of $C^{{\rm EE}}_\ell$  relative to the input power spectrum. For comparison, the green squares show the result of window-convolved power spectrum (see text for the details).}
\label{fig:validation_Emode}
\end{figure}
\begin{figure}[h]
\centering 
\includegraphics[width=0.45\textwidth]{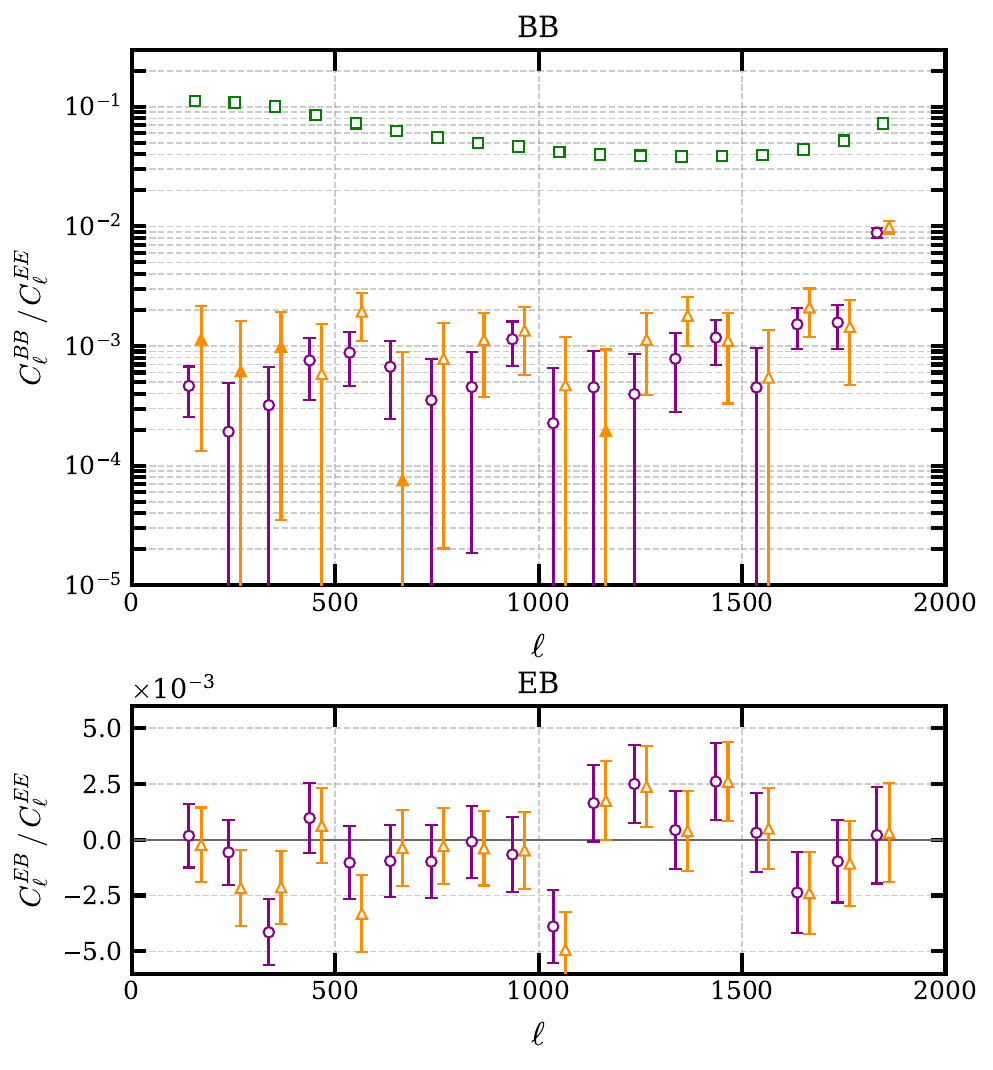}
\caption{
Ratio of the estimated power spectra, $C^{\mathrm{BB}}_\ell$ ({\it{Upper panel}}) and $C^{\mathrm{EB}}_\ell$ ({\it{Lower panel}}), to the input $E$-mode spectrum $C^{\mathrm{EE}}_{\mathrm{input}, \ell}$, using the Gaussian shear fields. 
The color coding and symbols follow Fig.~\ref{fig:validation_Emode}.
The $C^{\mathrm{BB}}_\ell$ values are plotted as absolute values, with filled (open) symbols denoting positive (negative) values.
}
\label{fig:validation_Bmode}
\end{figure}

The upper panel of Fig.~\ref{fig:validation_Emode} shows the $E$-mode angular power spectrum, $C^{\mathrm{EE}}_\ell$, measured using 20 linearly spaced bins across the multipole range $0 \le \ell \le 2000$.
As discussed in our previous paper~\citep{terawaki_quadratic_2025}, neither the optimal estimator nor the inverse-variance-weighted estimator can fully correct for power leakage from multipoles outside the range covered by the bins considered. Therefore, throughout this paper, we present results only for the 18 multipole bins excluding the two edge bins.
As shown in the bottom panel of Fig.~\ref{fig:validation_Emode}, both the optimal and inverse variance weighting schemes yield unbiased estimates of the true band power, maintaining accuracy to within a few percent.

We use 1,000 realizations for the ancillary vector fields to compute the Fisher matrix.
In this figure, we show only the results for an HSC-like shape noise level, corresponding to a source galaxy number density of $\bar{n}_s = 20\  \rm{arcmin}^{-2}$.
Error bars denote the $1\sigma$ standard error on the mean, computed from the standard deviation across 1,000 realizations.
For comparison, the green points show the window-convolved band powers corrected only for the masked area fraction.
They exhibit significant deviations from the underlying signal, highlighting the necessity of our estimator.
\begin{figure*}[t]
\centering 
\includegraphics[width=\textwidth]{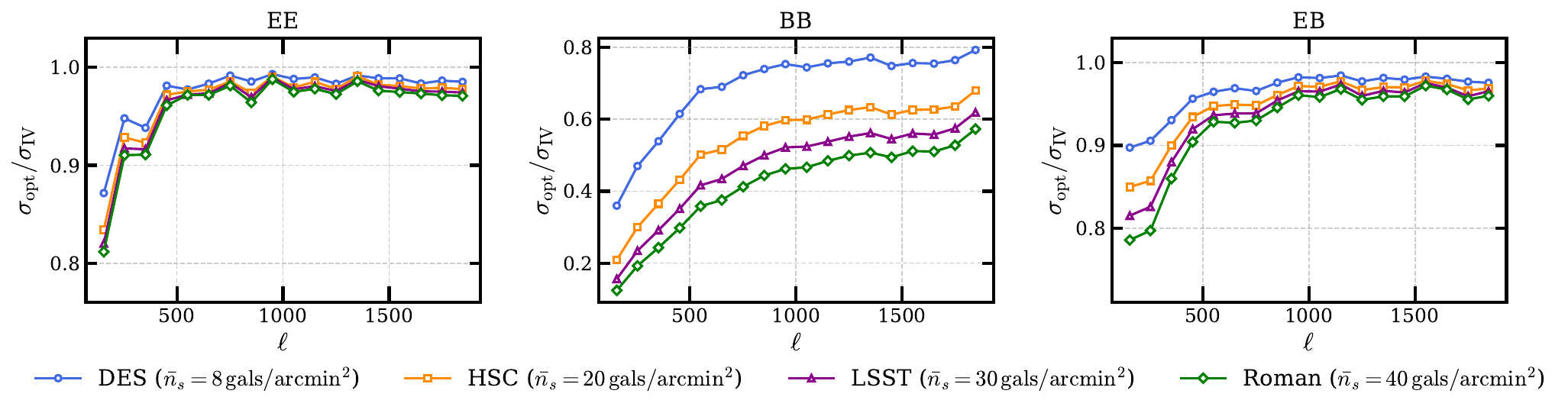}
\caption{
The ratio of the $1\sigma$ errors obtained with the optimal weighting scheme to those with the inverse variance (IV) weighting based on shape noise, calculated from 1,000 realizations of Gaussian shear fields. 
The ratios for $C^{\mathrm{EE}}_\ell$, $C^{\mathrm{BB}}_\ell$, and $C^{\mathrm{EB}}_\ell$ are presented from left to right. 
The blue circles, orange squares, purple triangles, and green squares show the results for the simulated data with different source galaxy number densities for shape noise:
DES ($\bar{n}_s=8\ \rm{arcmin}^{-2}$), HSC ($\bar{n}_s=20\ \rm{arcmin}^{-2}$), LSST ($\bar{n}_s=30\ \rm{arcmin}^{-2}$), and Roman ($\bar{n}_s=40 \ \rm{arcmin}^{-2}$), respectively.
Since we use the Gaussian shear fields and the true covariance matrix to estimate the Fisher matrix in the optimal estimator, this setup represents the theoretical limit of the achievable statistical precision.
}
\label{fig:error_ratio}
\end{figure*}

Fig.~\ref{fig:validation_Bmode} shows the measured $B$-mode
power spectra, $C^{\mathrm{BB}}_\ell$ and $C^{\mathrm{EB}}_\ell$. 
Although the complex survey window induces non-zero $C^{\mathrm{BB}}_\ell$, as evident from the convolved power in the upper panel, our estimator successfully deconvolves the window effect, effectively suppressing the leakage from $E$-modes into $B$-modes.
The lower panel of Fig.~\ref{fig:validation_Bmode} shows the measured $C^{\mathrm{EB}}_\ell$, which is consistent with zero within the error bars. 
Since $E$-to-$B$ leakage does not contribute to the $EB$ cross-spectrum because of parity symmetry, this result provides further validation of the robustness of our estimator.

The advantage of the optimal weighting scheme over inverse variance (IV) weighting is most pronounced in the sample-variance-dominated regime. 
Consequently, as the shape noise level decreases -- or equivalently, as the galaxy number density increases -- the gain achieved by the optimal weights becomes more significant. 
Fig~\ref{fig:error_ratio} compares the error bars obtained using the quadratic estimator for four different noise levels, corresponding to DES ($\bar{n}_s=8\ \text{arcmin}^{-2}$), HSC ($\bar{n}_s=20\ \text{arcmin}^{-2}$), LSST ($\bar{n}_s=30\ \text{arcmin}^{-2}$) and Roman ($\bar{n}_s=40 \ \text{arcmin}^{-2}$).
For $C^{\mathrm{EE}}_\ell$ ({\it{left}} panel), the reduction in error bars achieved by the optimal weighting scheme converges to a nearly constant level for $\ell > 500$, while it becomes significantly more pronounced at $\ell < 500$. 
This behavior is consistent with the theoretical expectation that the gain from optimal weighting is greatest in the sample-variance-dominated regime. 
This improvement also becomes more pronounced as the
noise level decreases: the gain ranges from 5--20\% for the DES-like case, increases to 10--30\% for the higher-density Roman configuration.
It is also worth noting that the optimal weights provide a modest improvement of a few percent, even in the shot-noise-dominated regime.

Similar to the $E$-mode results, the $B$-mode spectra ({\it{middle}} and {\it{right}} panel) exhibit a greater reduction in the error bars as the galaxy number density increases. 
Moreover, the reduction for $C_{\mathrm{BB}}$ is much larger than that for $C_{\mathrm{EE}}$, consistent with a previous study based on the full-sky simulation~\cite{maraio_testing_2023}. 
This suggests that our optimal weighting scheme is particularly advantageous for $B$-mode analyses, providing not only a more sensitive null test for systematic errors but also a powerful tool for probing the physical origin of $B$-modes,
including intrinsic alignments or other non-linear astrophysical processes.

\section{Application to non-Gaussian ray-tracing shear fields}
\label{sec:raytracing}

In this section, we perform a more realistic analysis using non-Gaussian shear fields. 
Although the optimality of the estimator is strictly guaranteed only under the assumption of Gaussian fields,
we investigate the extent to which the performance gain is preserved in the presence of non-Gaussianity and provide performance forecasts for various weak lensing surveys.

\subsection{Mock shear fields from ray-tracing simulations}
\label{subsec:raytracing}

To account for non-Gaussian effects, we utilize the full-sky ray-tracing simulations provided by \citet{takahashi_full-sky_2017}.\footnote{The simulation data are publicly available at \url{http://cosmo.phys.hirosaki-u.ac.jp/takahasi/allsky_raytracing/}.} 
These simulations consist of 108 light-cone realizations, each containing convergence and shear fields at multiple source redshifts in the \texttt{HEALPix} pixelization format. 
The ray-tracing simulations were generated from cosmological $N$-body simulation assuming a flat $\Lambda$CDM model consistent with the nine-year WMAP results~\citep{hinshaw_nine-year_2013}; specifically, the cosmological parameters are set to $\Omega_{\rm c}=0.233$, $\Omega_{\rm b}=0.046$, $\Omega_\Lambda=0.721$, $h=0.7$, $n_s=0.97$, and $\sigma_8=0.82$.
In this study, we use the simulation data provided on a \texttt{HEALPix} pixel with $N_{\rm side}=8192$, corresponding to an angular pixel size of $0.43$~arcmin.
Although \citet{takahashi_full-sky_2017} provides full-sky simulations, we adopt the flat-sky approximation so that the shear field can be analyzed using the FFT. 
By properly accounting for the projection of the spin-2 spherical-harmonic basis, we project the shear field on the celestial sphere onto the flat sky to construct flat-sky shear maps (see Terawaki et al., in preparation, for details).
In this step, we map the \texttt{HEALPix} data onto 2D flat Cartesian pixels using the Nearest Grid Point (NGP) method~\citep{sefusatti_accurate_2016}. 
Following this procedure, we extract 10 non-overlapping regions along the equator from each of the 108 realizations, resulting in a total of 1,080 flat-sky shear map realizations.
Each region covers an area of $\ang{20}\times\ang{20}$ and 
we use a $500\times 500$ pixels for the FFT.
We then apply the 15\% star masks and zero-padding around the data region, following the procedure described in 
Section~\ref{sec:validation}, resulting in a total FFT area of $\ang{24}\times\ang{24}$ and a $600\times 600$ grids.

To mimic realistic weak lensing data, we simulate the ``observed'' distortion field at each pixel by combining the convergence $\kappa$ and shear fields ($\gamma_1, \gamma_2$) from a single redshift plane at $z=1.03$. 
Following Eqs.~(24) and (25) of \citet{2019MNRAS.486...52S}, we add intrinsic shape noise, generated in the same manner as described in Section~\ref{sec:validation}, to the simulated lensing signals. 
Because we model the distortion field directly, we incorporate the shear responsivity \citep{2002AJ....123..583B} when estimating the shear field from the distortion fields.

\subsection{Performance with non-Gaussian shear fields
}
\label{subsec:nongauss_result}
\begin{figure}[h]
\centering 
\includegraphics[width=0.45\textwidth]{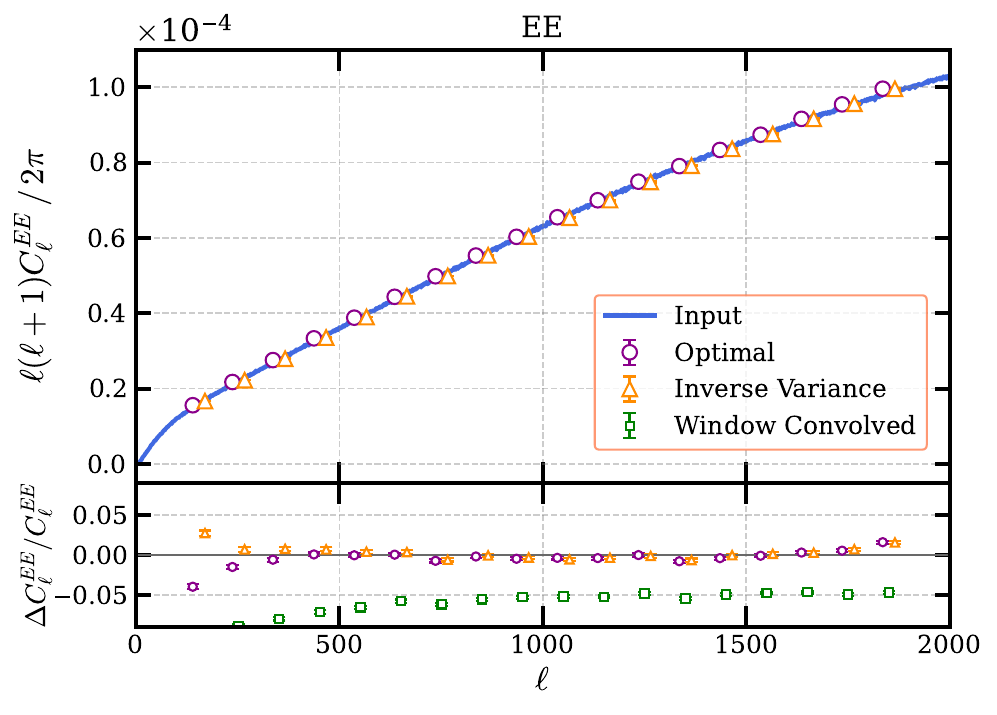}
\caption{Similar to Fig.~\ref{fig:validation_Emode}, but using
galaxy shape fields generated from the non-Gaussian ray-tracing simulations of a $\Lambda$CDM cosmology (see text for details).
An HSC-like shape noise contribution is incorporated into the simulated shear fields. 
Each simulated shape field covers an area of $576\ \text{deg}^2$ and includes star masks and zero-padding regions surrounding the data region, as in Fig.~\ref{fig:window}, under the flat-sky approximation.
We also plot the ``input'' power spectrum (blue line),
computed from the same simulations in their original full-sky form without shape noise or a survey window function.
The color and plotting schemes are identical to those in Fig.~\ref{fig:validation_Emode}. 
Each data point represents the mean band power averaged over 1,080 realizations, with error bars denoting the $1\sigma$ uncertainty on the mean.
}
\label{fig:validation_Emode_rt}
\end{figure}

\begin{figure}[h]
\centering 
\includegraphics[width=0.45\textwidth]{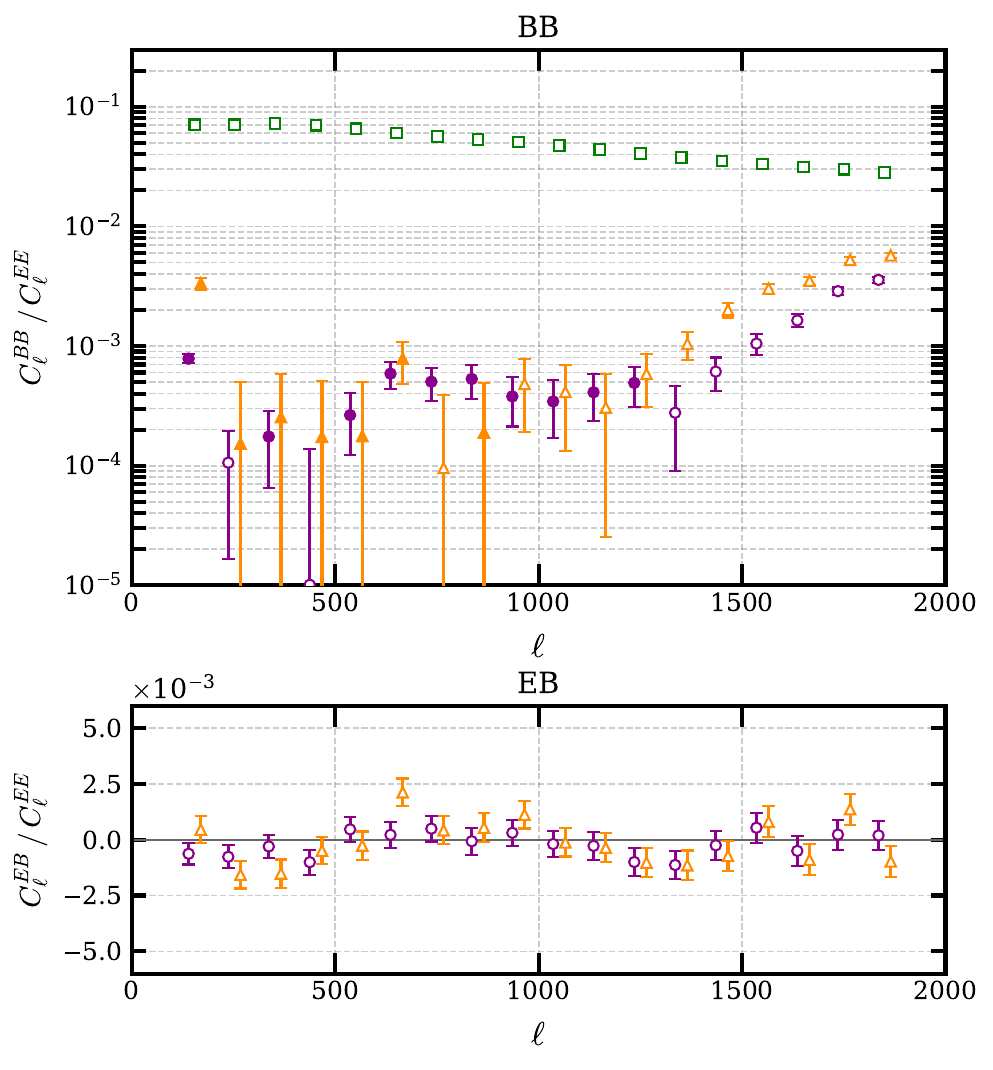}
\caption{Similar to Fig.~\ref{fig:validation_Bmode}, but using the simulated shape fields 
generated from the ray-tracing simulations of a $\Lambda$CDM model as in Fig.~\ref{fig:validation_Emode_rt}.
}
\label{fig:validation_Bmode_rt}
\end{figure}

\begin{figure*}[t]
\centering 
\includegraphics[width=\textwidth]{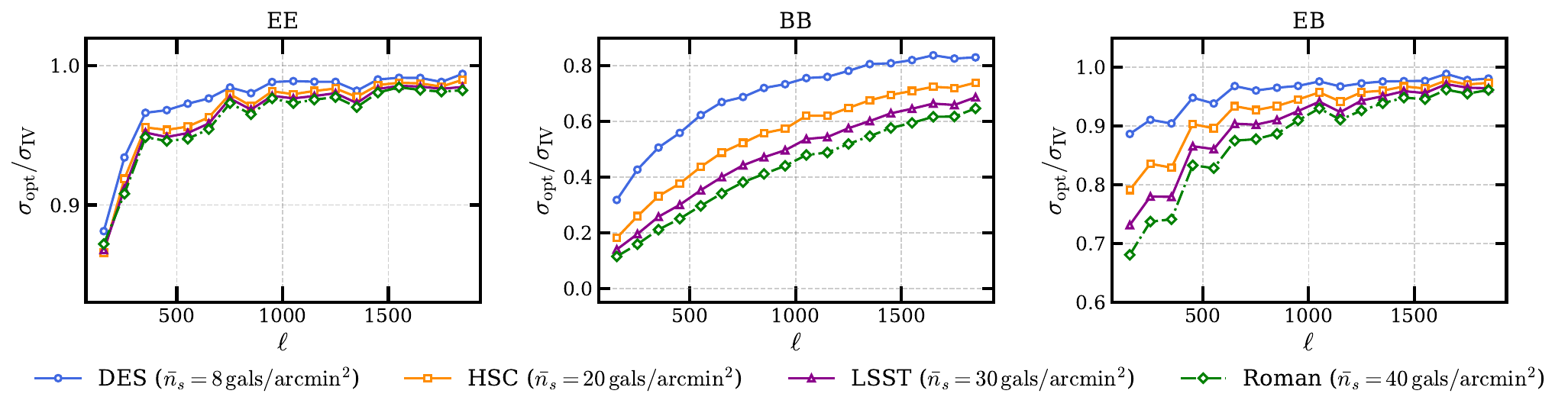}
\caption{Similar to Fig.~\ref{fig:error_ratio}, but using the simulated shape fields generated from the ray-tracing simulations of a $\Lambda$CDM model as in Fig.~\ref{fig:validation_Emode_rt}.
}
\label{fig:error_ratio_rt}
\end{figure*}

Figs.~\ref{fig:validation_Emode_rt} and \ref{fig:validation_Bmode_rt} present the cosmic shear power spectra measured using the quadratic estimator. 
For comparison, the ``true'' power spectrum is computed from the full-sky simulations without adding shape noise or applying the survey window function. 
In both plots, the color and plotting schemes are identical to those in Figs.~\ref{fig:validation_Emode} and \ref{fig:validation_Bmode}. 
Note that we subtracted the residual shape-noise contribution from the estimated power spectrum to improve its agreement with the true power spectrum at high multipoles.
We divide the measured power spectra by the pixel window function to correct for pixelization effect.
The error bars represent the $1\sigma$ uncertainty on the mean.
As shown in the bottom panel of Fig.~\ref{fig:validation_Emode_rt}, both optimal and inverse-variance weighting schemes recover the true band powers within a few percent.

Fig.~\ref{fig:error_ratio_rt} shows the extent to which the quadratic estimator reduces the statistical errors relative to the conventional suboptimal estimator.
Compared to Fig.~\ref{fig:error_ratio}, the error reduction is smaller because our estimator is strictly optimal only for Gaussian fields.
Nevertheless, we still find an improvement of approximately 10\% on large scales and a few percent reduction on small scales for the $E$-mode.
These improvements become more significant for observations with higher galaxy number densities. 
As in the Gaussian case, the error bars of the $C^{\rm{BB}}_\ell$ measurement are reduced much more significantly than those of the $C^{\rm{EE}}_\ell$ and $C^{\rm{EB}}_\ell$.

\subsection{Joint estimation of $E/B$-mode power spectra}
\label{sec:EB_joint}
\begin{figure}[h]
\centering 
\includegraphics[width=0.45\textwidth]{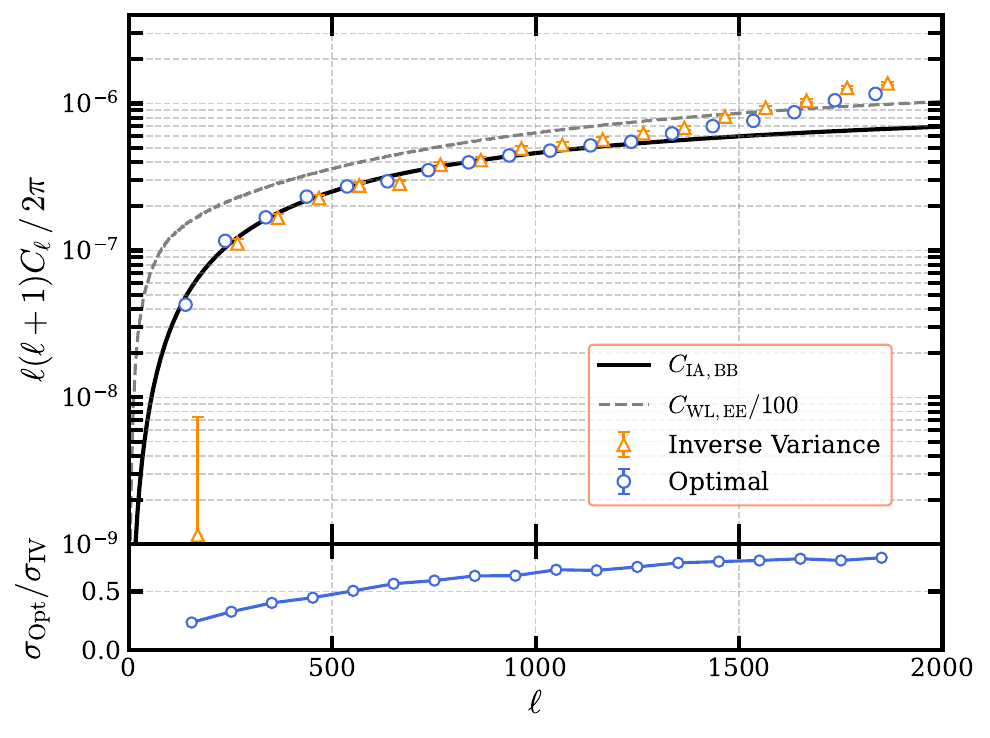}
\caption{{\it{Upper panel}}: The $B$-mode power spectrum $C^{\rm BB}_\ell$ estimated using the optimal estimator and the inverse variance (IA) weighting for the simulated shape fields with a non-zero input $B$-mode power spectrum. 
We add $B$-mode fields predicted from the intrinsic alignment (IA) model under the Gaussian-field assumption to the shape fields generated from the ray-tracing simulations as in Fig.~\ref{fig:validation_Emode_rt}.
The amplitude of the IA $B$-mode power spectrum is set to approximately 1/100 of that of the cosmic shear $E$-mode  power spectrum, as indicated by the dashed line, while its shape follows the prediction of the TATT model (see the text for details).
{\it{Lower panel}}: The ratio of the 1$\sigma$ errors, similar to Fig.~\ref{fig:error_ratio}.
}
\label{fig:IABmode}
\end{figure}

In the standard $\Lambda$CDM model, under the weak-lensing approximation, scalar gravitational potentials generate only $E$-mode shear fields, with no $B$-mode component.
In real observations, however, observational systematic effects, such as imperfect PSF correction, can generate a non-vanishing $B$-mode signal. 
In addition, nonlinear effects, 
such as intrinsic alignments (IA), generally generate both $E$- and $B$-mode fields. 
In this section, we consider simulated galaxy-shape fields containing both $E$ and $B$ modes and examine whether the quadratic estimator can jointly recover their power spectra.
To generate physically motivated $B$-modes, we adopt the Tidal Alignment and Tidal Torquing (TATT) model \citep{2019PhRvD.100j3506B} for intrinsic alignments. 
We then evaluate the performance of our proposed estimator in measuring the $B$-mode power spectrum.

We generate Gaussian $B$-mode fields based on the predictions of the TATT model and then add them to the galaxy shape fields generated from the ray-tracing simulations used in Fig.~\ref{fig:validation_Emode_rt}.
We use the \texttt{CCL} software to model the $B$-mode power spectrum based on the TATT model, adopting the parameter values $A_1=0.46, A_2=-0.37$, and $A_{1\delta}=0.161$.
Here $A_1$, $A_2$ and $A_{1\delta}$ correspond to $A_1$, $A_2$ and $b_{\rm ta}$, respectively, in Eq.~(20) of Ref.~\cite{2023PhRvD.108l3518L}, with $A_{1\delta}=b_{\rm ta}A_1$. 
These values are taken from the best-fit values of Ref.~\cite{2023PhRvD.108l3518L}.
Although the amplitude of the IA-induced $B$-mode power spectrum depends on the redshift distribution of the galaxies responsible for the intrinsic alignments, we set its amplitude to be approximately 1/100 of that of the cosmic shear $E$-mode power spectrum.

Fig.~\ref{fig:IABmode} shows the $B$-mode power spectra estimated using the optimal estimator or the pseudo-$C_\ell$ estimator, as in Fig.~\ref{fig:validation_Bmode_rt}. 
We again note that, as in Fig.~\ref{fig:validation_Emode_rt},
we subtracted the residual shape-noise contribution from the estimated $B$-mode spectrum.
The figure shows that the optimal estimator more accurately recovers the input $B$-mode power spectrum than the pseudo-$C_\ell$ estimator. This demonstrates that our method enables more stringent tests for a non-zero $B$-mode power spectrum by minimizing the leakage of $E$-mode power into the $B$-mode spectrum.
We note that the improvement in the error bars for the non-zero $B$-mode spectrum is similar to that shown in the middle panel of Fig.~\ref{fig:error_ratio_rt}.

\section{Conclusions}
\label{sec:conclusion}

In this paper, we have developed an optimal estimator for the ``unwindowed'' $E$- and $B$-mode cosmic shear power spectra by maximizing the Gaussian likelihood function for the galaxy shape field (see Section~\ref{sec:method}). 
The estimator requires estimating the Fisher matrix, which involves repeated inverse-covariance matrix operations. 
By combining the flat-sky approximation, the conjugate-gradient method, and Monte Carlo realizations of Gaussian ancillary fields, we made the Fisher matrix estimation computationally tractable.
Using Gaussian simulations of galaxy shape fields, we demonstrated that the estimator accurately recovers the underlying $E$-mode power spectrum with optimal statistical precision across all angular scales (Figs.~\ref{fig:validation_Emode}--\ref{fig:error_ratio}).

Using galaxy shape fields generated from ray-tracing simulations of a $\Lambda$CDM cosmology, we showed that the method improves the statistical precision of the recovered 
$E$-mode power spectrum by up to 15\% in low-multipole bins within the sample-variance-dominated regime, compared with the pseudo-$C_\ell$ estimator~(Figs.~\ref{fig:validation_Emode_rt}--\ref{fig:error_ratio_rt}).
The $15\%$ improvement is significant as it is equivalent to increasing the survey area by approximately 30\%.
The improvement becomes more significant for cosmic shear surveys with higher number density of source galaxies, which extend the range of angular scales over which sample variance dominates.
We also showed that the method can suppresses the leakage of $E$-mode power into the $B$-mode power spectrum~(Fig.~\ref{fig:validation_Bmode_rt}).
We also discussed how the improved $B$-mode power spectrum enables more sensitive tests for the presence of a non-zero $B$-mode power spectrum, such as that induced by intrinsic alignments~(Fig.~\ref{fig:IABmode}).

Since the standard 2PCF estimator typically employs inverse-variance weighting based on the shape noise, our method is expected to yield power spectrum measurements with higher signal-to-noise ratios, particularly on large angular scales, 
from the same survey data.
The improved measurements of $E$-mode cosmic shear power spectrum will enable us to extract more cosmological information.
We plan to apply the method to existing cosmic shear surveys, such as the Subaru HSC survey, to derive cosmological constraints. 
Our method will also be valuable for upcoming next-generation surveys, including LSST and the Nancy Grace Roman Space Telescope.

Another promising direction is to extend the method to the estimation of the cosmic shear bispectrum. 
In this case, the estimator must simultaneously recover all eight bispectra corresponding to different combinations of the $E$- and $B$-mode shear fields (e.g., $EEE$, $EBB$, and $BBB$) for arbitrary triangle configurations. 
Since the power spectrum and bispectrum have different dependence on redshift, scale, and $E$/$B$-modes, combined measurements of the cosmic shear power spectra and bispectra will not only improve constraints on cosmological parameters but also provide more stringent tests of systematic effects. 
Extending our method to enable such a joint analysis is an important direction for future work.

\acknowledgments
We would like to thank 
Ryo~Terasawa, 
Oliver~Philcox,
Adri~Duivenvoorden, 
Mat~Madhavacheril, 
Yuji~Chinone and Akito~Kusaka for very useful discussion.
This work was supported in part by
JSPS KAKENHI Grant Number 
24H00215, 26H00401,
26H00404,
and by World Premier International Research Center Initiative (WPI Initiative), MEXT, Japan.

\bibliography{refs}

\end{document}